\documentclass{emulateapj}
\usepackage{graphicx}

\slugcomment{To appear in Astrophysical Journal Letters}
\shorttitle{Photoevaporation-starved Accretion}
\shortauthors{J.J.~Drake et al.}
\begin{document}

\title{X-ray Photoevaporation-starved T~Tauri Accretion} 

\author{Jeremy J. Drake\altaffilmark{1},
Barbara Ercolano\altaffilmark{2},
Ettore Flaccomio\altaffilmark{3},
Giusi Micela\altaffilmark{3}
}

\affil{$^1$Smithsonian Astrophysical Observatory,
MS-3, 60 Garden Street, Cambridge, MA 02138}
\email{jdrake@cfa.harvard.edu}
\affil{$^2$Institute of Astronomy, University of Cambridge, Madingley Road, 
Cambridge, CB3 0HA,  UK}
\affil{$^3$INAF Osservatorio Astronomico di Palermo, Piazza del Parlamento 1,
90134 Palermo, Italy}

\begin{abstract}
X-ray luminosities of accreting T~Tauri stars are observed to be systematically lower than those of non-accretors.  There is as yet no widely accepted physical explanation for this effect, though it has been suggested that accretion somehow suppresses, disrupts or obscures coronal X-ray activity.  Here, we suggest that the opposite might be the case: coronal X-rays modulate the accretion flow.  We re-examine the X-ray luminosities of T~Tauri stars in the Orion Nebula Cluster and find that not only are accreting stars systematically fainter, but that there is a correlation between mass accretion rate and stellar X-ray luminosity.  We use the X-ray heated accretion disk models of Ercolano et al.\ to show that protoplanetary disk photoevaporative mass loss rates are strongly dependent on stellar X-ray luminosity and sufficiently high to be competitive with accretion rates.  X-ray disk heating appears to offer a viable mechanism for modulating the gas accretion flow and could be at least partially responsible for the observed correlation between accretion rates and X-ray luminosities of T~Tauri stars.
\end{abstract}


\maketitle

\section{Introduction}
\label{s:intro}


A survey of the X-ray emission from T~Tauri stars in the Taurus-Auriga star-forming region by \citet{Neuhaeuser.etal:95} based on ROSAT observations revealed a significant difference amounting to a factor of 2-3 between the X-ray luminosities of classical T~Tauri stars (CTTS) still in the phase of active accretion, and weak T~Tauri stars (WTTS) of the same mass that exhibit no accretion signatures.  This finding was confirmed in subsequent ROSAT studies of Taurus-Auriga and other nearby star-forming regions \citep[e.g.][]{Alcala.etal:97,Wichmann.etal:97,Stelzer.Neuhauser:01, Flaccomio.etal:03b}, in later {\it XMM-Newton} surveys of the Taurus molecular cloud \citep{Telleschi.etal:07} and NGC~2264 \citep{Dahm.etal:07}, and by {\it Chandra} X-ray surveys of low-mass stars in the Orion Nebula Cluster (ONC) \citep{Flaccomio.etal:03,Stassun.etal:04,Preibisch.etal:05}, and NGC~2264 \citep{Flaccomio.etal:06}.
This general result might be considered surprising, since the additional energy input through accretion might naively be expected to result in {\it additional} X-ray emission through ballistic accretion shocks rather than in diminished X-rays \citep[e.g.][]{Lamzin.etal:96,Lamzin:99,Calvet.Gullbring:98}.

Several different explanations for the X-ray deficiency of accreting
T~Tauri stars have been proposed 
\citep[e.g][]{Flaccomio.etal:03,Stassun.etal:04,Preibisch.etal:05,
  Jardine.etal:06,Gudel.etal:07b,Gregory.etal:07}, including: disruption of the magnetic corona by the encroaching accretion disk; supression of convection through the influence of accreted gas on stellar structure; reduction in differential rotation due to star-disk interactions; and obscuration of X-ray emission by accretion funnels.  All of these explanations involve the suppression of X-rays as a consequence of accretion.  We suggest here that, instead of accretion modulating X-ray emission, X-ray emission itself might be a driver modulating the accretion.

Recent results of two-dimensional photoionisation and radiative transfer
calculations of the heating of protoplanetary disks by parent stellar
coronal X-rays by \citet{Ercolano.etal:08b} indicate that the X-rays
drive a disk coronal wind of significant mass flux.  The estimated
mass loss rate for a $0.75~M_\odot$ star with a rather typical ONC
X-ray luminosity of $2\times 10^{30}$~erg~s$^{-1}$ was of order 
$10^{-8}\ M_\odot$~yr$^{-1}$.  Moreover, the mass loss was dominated by
the region of the disk lying at a radial distance of 10-20~AU.  In
principle, such a photoevaporation zone could drive off a significant
fraction of the migrating gas from the outer disk and 
starve the inner disk of gas to replenish the accretion flow onto 
the star.  The photoevaporation rate, and the extent to which inflowing
gas is driven off the disk, should be correlated with the 
stellar X-ray luminosity.

Here, we present new evidence that not only are the accretors less
X-ray luminous than non-accretors, but that the mass accretion rates
of ONC CTTS are inversely correlated with X-ray luminosity, as also
suggested recently for Taurus molecular cloud stars by
\citet{Telleschi.etal:07}.  We propose that this could be a signature
of X-ray photoevaporation-starved accretion flow: stars that are more
X-ray bright are naturally accreting at lower rates because the supply
of gas to the inner disk is diminished by photoevaporative dissipation.  In
\S\ref{s:photevap} we outline extentions to the disk model
calculations of \citet{Ercolano.etal:08b} to investigate the influence
of X-ray luminosity on photoevaporation rate, and in \S\ref{s:massacc}
we present a fresh analysis of the Ca~II~8662~\AA\ accretion
diagnostic and stellar X-ray luminosities for stars of the ONC
observed in the {\it Chandra Orion Ultra-Deep survey} 
(COUP; \citealt{Feigelson.etal:05}.
The results are discussed and summarised in
\S\ref{s:discuss} and \$\ref{s:summary}.

\section{X-ray photoevaporation of disk gas}
\label{s:photevap}

\citet{Ercolano.etal:08b} have recently presented an estimate of the photoevaporation rate of gas in a protoplanetary disk driven by X-ray heating.  
In order to gain further understanding of the dependence of the mass loss rate on X-ray luminosity, we have performed photoevaporation rate calculations for identical disk parameters used by \citet{Ercolano.etal:08b},  but for different values of the 
stellar X-ray luminosity.  

The basis of our X-ray photoevaporation estimate is a 2D photoionisation and dust radiative transfer model calculated using the {\sc mocassin} Monte Carlo code \citep{Ercolano.etal:03c,Ercolano.etal:05,Ercolano.etal:08c} of a prototypical T~Tauri disk irradiated by X-rays.  The model of \citet{Ercolano.etal:08b} assumed a 0.1-10~keV luminosity for the central pre-main sequence star of $L_X=2\times 10^{30}$~erg~s$^{-1}$; here we performed analogous calculations for values of X-ray luminosity ranging from $2\times 10^{29}$ to $4\times 10^{31}$~erg~s$^{-1}$, bracketing the value adopted by 
\citet{Ercolano.etal:08b} by  factors of 10.  The X-ray source was assumed to be an optically-thin, collision-dominated plasma with a temperature of $\log T=7.2$.  The disk was assumed to have the initial dust and gas density distribution of a 2D disk model computed by \citet{D'Alessio.etal:98} for a central star of mass 0.7~M$_\odot$, 
radius 2.5~R$_\odot$, and effective temperature 4000~K, and for 
an accretion rate $\dot M=10^{-8}$~M$_\odot$~yr$^{-1}$,  viscosity parameter $\alpha=0.01$ (see, e.g., \citealt{Pringle:81}), and total disk mass 0.027~M$_\odot$.   
In a 
refinement over the method used by  \citet{Ercolano.etal:08b}, the vertical gas density structure was additionally iterated from the starting solution to preserve hydrostatic equilibrium (HSE) using a version of the formalism of \citet{Alexander.etal:04} , modified to account for significant vertical extension of the disk gas.
A more complete description 
of the model calculations is presented by \citet{Ercolano.etal:09}.

Following \citet{Alexander.etal:04} and \citet{Ercolano.etal:08b}, 
the photoevaporation rate through
X-ray heating at each radial grid point, $R$, was estimated by
assuming the base of the flow is located at the height where the local
gas temperature exceeds the escape temperature of the gas, defined as
$T_{es} = G m_H M_* / k R$, where $M_*$ is the stellar mass.  To a
first approximation, the mass loss rate per unit area is $\rho c_s$,
where $\rho$ and $c_s$ are the gas density and the sound speed
evaluated at the base of the flow, respectively.

\begin{figure}
\plotone{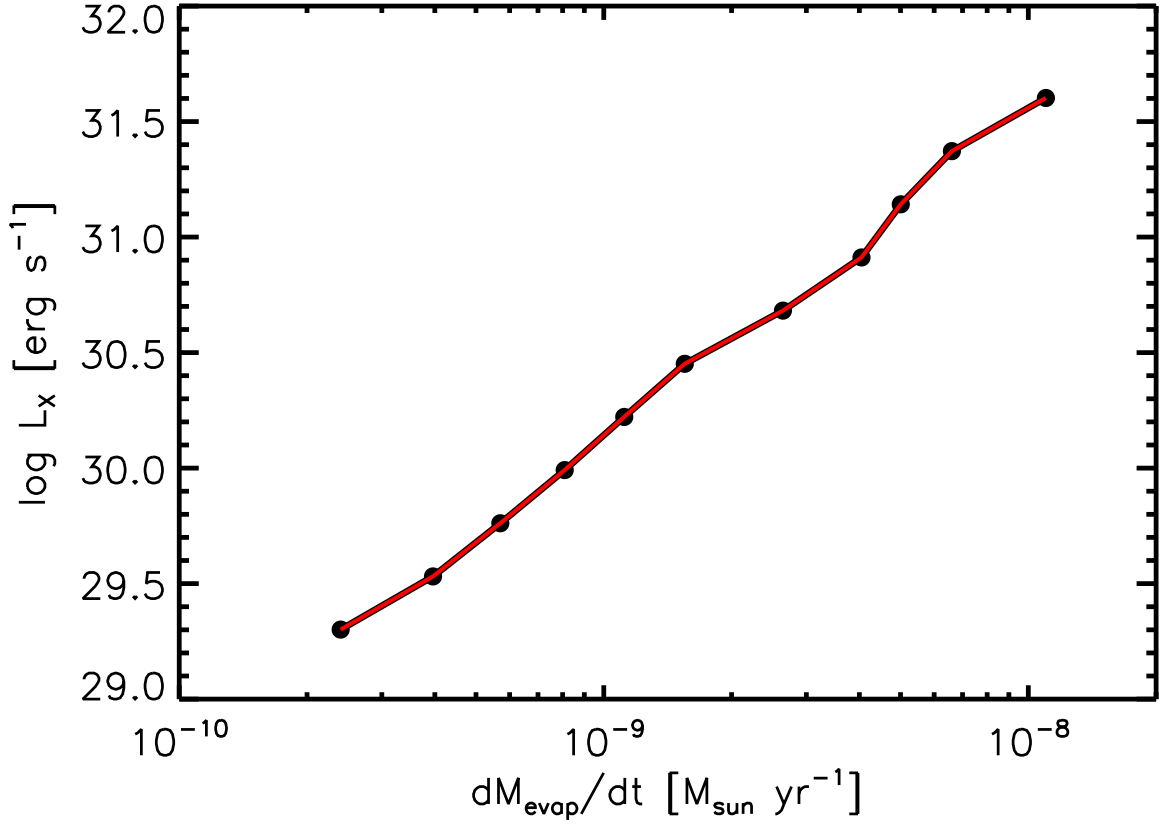}
\caption{
X-ray heating-induced photoevaporation rate, $\dot{M}$, from a 2D 
protoplanetary disk model expressed 
as a function of the model 
coronal X-ray luminosity of the central star, $L_X$, computed using the 
{\sc  Mocassin} program (see text and \citealt{Ercolano.etal:08b}
for details).
}
\label{f:mdot_lx}
\end{figure}

\begin{figure}
\plotone{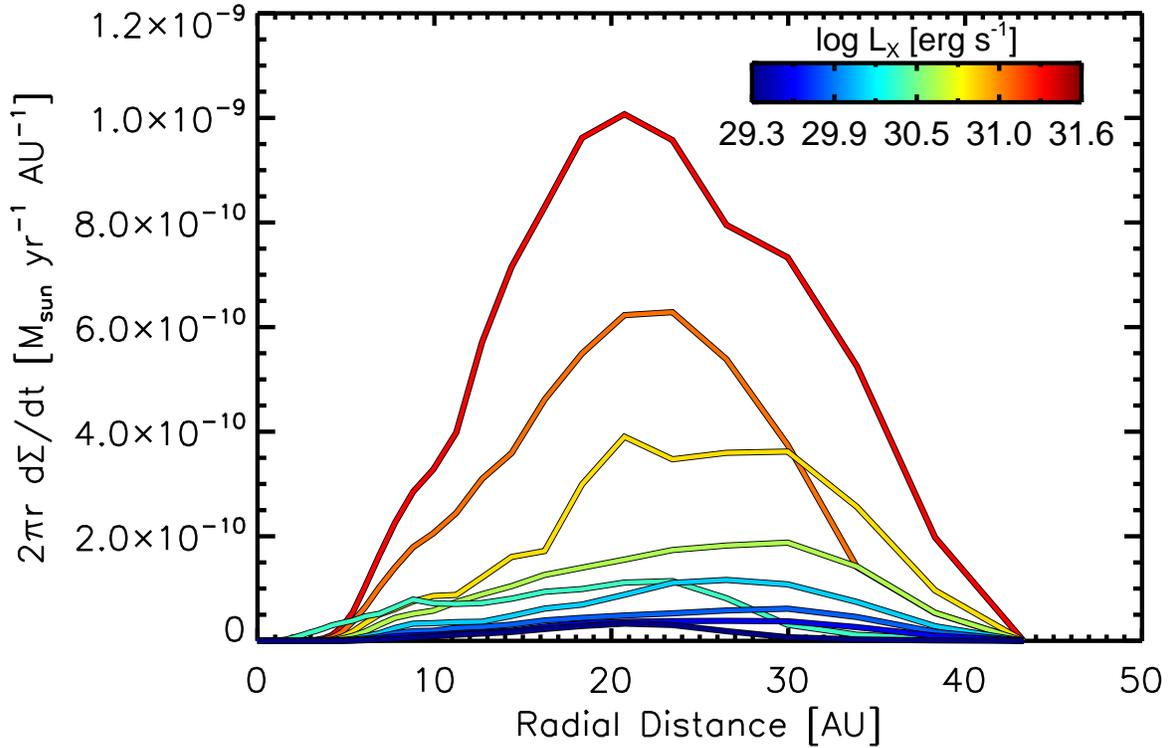}
\caption{
The photoevaporation rate per unit radius, $d\Sigma/dt$, as a function of 
radial distance from the 
central star for different stellar X-ray luminosities, $L_X$.   The evaporation rate 
peaks in the region 20~AU and at high values of $L_X$ can compete with
the inward viscous flow, starving the inner disk of the replenishing
gas from the outer disk that contains the bulk ($\sim 90$\% ) of
the total disk gas mass.
}
\label{f:mdot_rad}
\end{figure}

The total disk mass loss rate as a function of stellar X-ray
luminosity is illustrated in Figure~\ref{f:mdot_lx}, and the mass loss
rate integrated over an annulus of unit width as a function of the
annular radius, $R$, is illustrated for models corresponding to different 
stellar X-ray luminosities in Figure~\ref{f:mdot_rad}.  As noted by
\citet{Ercolano.etal:08b}, the mass loss rate peaks strongly in the
10-20~AU range---a zone far enough from the central star that escape
temperatures become sufficiently low for the decreasing X-ray heating
with increasing radius to effect the maximum mass loss.  The mass loss
rate is found to depend strongly on stellar X-ray luminosity, as
expected.

The photoevaporation rate for the canonical model with $L_X=2\times 10^{30}$~erg~s$^
{-1}$ is about a factor of ten lower than the non-HSE, fixed density 
structure calculation of \cite{Ercolano.etal:08b}: this is due to the screening capacity of 
the vertical evaporated gas flow that acts to partially shield the disk from X-rays.  The 
HSE calculations here will in fact overestimate this screening because the true gas 
density in outflowing unbound gas will be lower than that for the HSE solution in which 
the gas pressure has to support the overlying column.  The HSE photoevaporation rate 
will then underestimate the true photoevaporation rate.  Nevertheless, within the bounds 
of accuracy of the calculations, we find X-ray photoevaporation rates comparable with 
observed accretion rates of CTTS. 
We also find a strong dependence of 
photoevaporation rate on $L_X$: for a factor of 100 change in $L_X$
the disk mass loss rate changes by a factor of $\sim 50$.

\section{Ca~II and an $L_X$--$\dot M$ relationship for the ONC}
\label{s:massacc}



Strong blue and near infrared lines of Ca~II are formed purely in absorption in nearly all late-type stellar photospheres, with some sensitivity to stellar effective temperature and metallicity \citep[e.g.][]{Smith.Drake:87,Smith.Drake:90,Chmielewski:00}.  In magnetically active stars, the lines are observed to be filled in by emission from the chromosphere.  In the case of accreting T~Tauri stars, the photospheric line in-filling is thought to be caused also by emission from regions of the upper atmosphere heated by magnetospheric accretion.  This occurs to such an extent that the lines can be seen in emission, and the emission line strength of the infrared Ca~II lines has been found an excellent diagnostic of mass accretion rate, $\dot M$, \citep[e.g.][]{Batalha.etal:96,Muzerolle.etal:98}.

\citet{Flaccomio.etal:03} used the Ca~II~8662~\AA\ equivalent width measurements of \citet{Hillenbrand.etal:98} to divide their {\it Chandra} sample of X-ray detected T~Tauri stars in the Orion Nebular Cluster (ONC) into ``high'' and ``low'' accretion rates, and then to find differences in the X-ray luminosity functions of these two groups \citep[see also][]{Stassun.etal:04,Preibisch.etal:05}.  Here, we re-visit the \citet{Hillenbrand.etal:98} Ca~II~8662~\AA\ equivalent width data with a view to making a more precise assessment of the mass accretion rates of ONC stars for which both Ca~II and X-ray luminosity measurements exist.

The relationship between the observed Ca~II~8662~\AA\ flux,
$F_{CaII}$, and mass accretion rate, $\dot M$, has been studied in
detail by \citet{Mohanty.etal:05} for a wide range of stellar mass.
For CTTS, they find (their Eqn.~2)
\begin{equation}
\dot{M}=0.71 \log_{10}(F_{CaII}) -12.66.
\label{e:mdot_fcaii}
\end{equation}
We adopt this relationship as the basis of our $\dot{M}$ calibration for the ONC sample,
together with the compilation of relevant data from  \citet{Getman.etal:05}.  The 
latter includes X-ray luminosities from the COUP survey, Ca~II equivalent widths from 
\citet{Hillenbrand.etal:98}, stellar effective temperatures and bolometric luminosities updated from \citet{Hillenbrand:97}, and stellar masses inferred from the evolutionary tracks of \citet{Siess.etal:00}. 

Since Eqn.~\ref{e:mdot_fcaii} involves the {\em
  flux} of the Ca~II~8662~\AA\ line, rather than the equivalent width, we have converted the latter measurements of \citet{Hillenbrand.etal:98} into fluxes using the predicted continuum fluxes of the model atmospheres of \citet{Allard.etal:00}.  These models span an effective temperature range of 3000-5000~K, and the logarithmic surface gravity range 3.5-5.5.  The 8662~\AA\ continuum intensity for each star was obtained by interpolation in the model grid for the appropriate values of effective temperature and surface gravity.  We derived the latter from the masses, temperatures and luminosities tabulated by \citet{Getman.etal:05}.  Out of 502 stars with spectral types, effective temperatures and Ca~II equivalent width measurements, we obtained $\dot{M}$ estimates for 412.  We further restricted the sample range to stars with mass $0.5M_\odot<M<2.0M_\odot$: being intrinsically more faint, stars of lower mass are prone to larger uncertainties in observed X-ray fluxes and samples are inherently less complete; higher mass stars have essentially radiative envelopes and likely exhibit quite different X-ray and disk behaviour.  The stars of this sample with X-ray luminosity measurements 
numbered 120, and of these 45 have mass accretion rates $\dot M> 0$; the remainder 
have Ca~II~8662~\AA\ either in absorption or with an equivalent width of zero.

\begin{figure}
\plotone{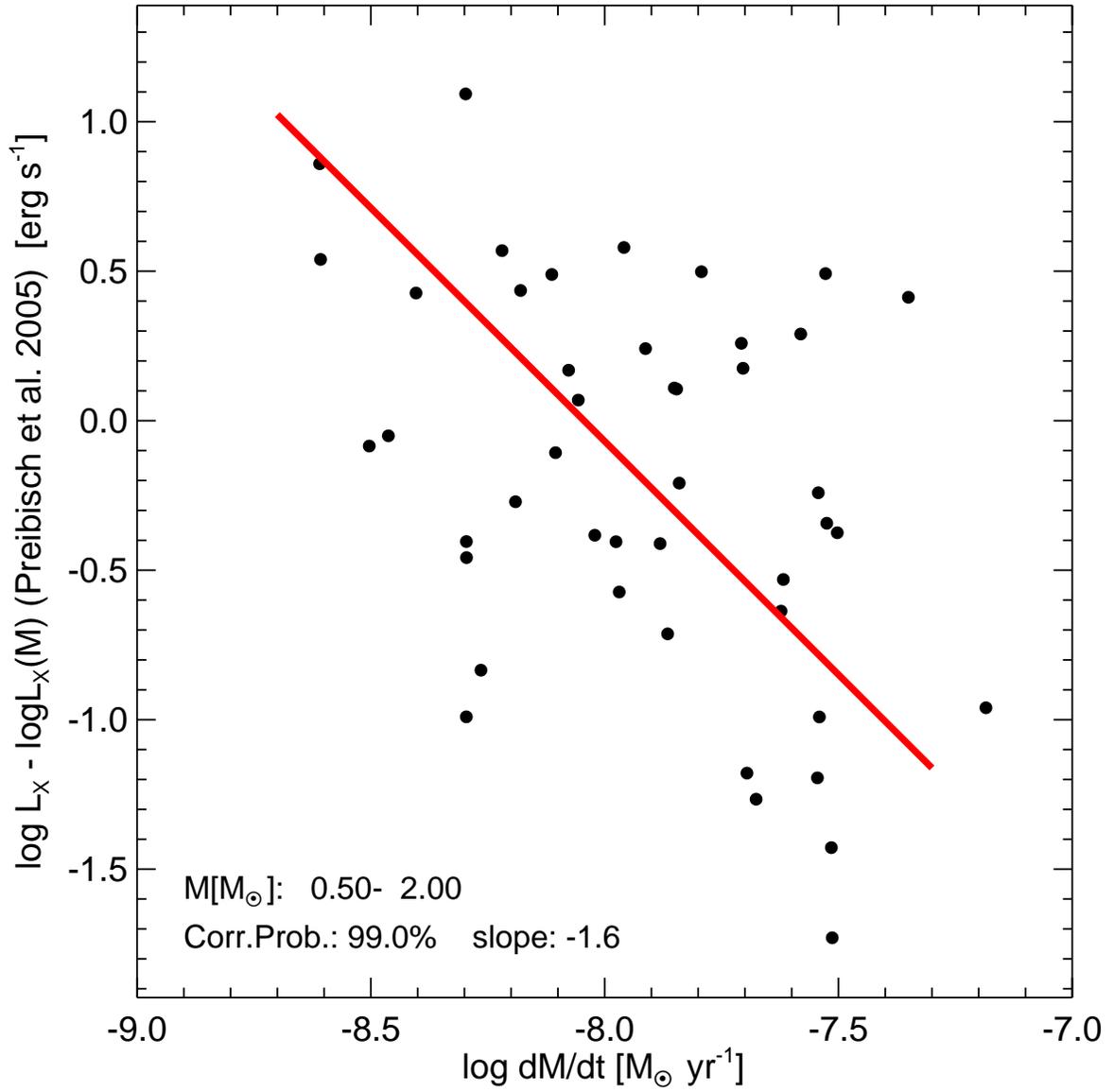}
\caption{
The deviation of ONC T~Tauri X-ray luminosity from the mean $L_x$ vs.\
stellar mass, $M$, relation of \citet{Preibisch.etal:05} as a function
of derived mass accretion rate (see text).  The Spearman's $\rho$ correlation
probability is 99\%.  The correlation indicates that, on average, stars with 
higher X-ray luminosity have lower mass accretion rate, $\dot{M}$. 
}
\label{f:mdot_lxdev}
\end{figure}

Since the final sample of 45 stars covers a significant mass range, and pre-main sequence stellar X-ray luminosity depends on stellar mass, we studied the $L_X$--$\dot M$ relationship by comparing our derived mass accretion rates with {\em deviations} from the mean $L_X$~vs~$M$ relation derived for the COUP sample by \citet{Preibisch.etal:05}, $\Delta \log L_X= \log L_X -\log L_X(M)$.  This is illustrated in Figure~\ref{f:mdot_lxdev}, where we find evidence for an {\em inverse} correlation between $\dot M$ and $\Delta \log L_X$: the Spearman's $\rho$ correlation probability is 99\%\ and the best-fit slope is $-1.6$.  Similar anti-correlations are seen in  
$L_{X}/L_{bol}$ vs $\dot M$ 
(correlation probability 99.94\%, slope$=-1.9$) and in $L_{X}/L_{bol}$ vs $\dot M/M$ (correlation probability 99.96\%, slope$=-1.6$).
These data provide strong evidence that stars of a given mass with higher X-ray luminosity have lower mass accretion rates: the initial finding of \citet{Neuhaeuser.etal:95} that WTTS are generally more X-ray bright than CTTS appears to be the manifestation of a continuous relation between X-ray luminosity and accretion rate.

\section{Discussion}
\label{s:discuss}

In \S\ref{s:intro}, it was noted that current ideas attempting to explain the lower X-ray luminosities of accreting T~Tauri stars all seek to exploit possible suppression or disruption of coronal activity by accretion.  Since the X-ray photoevaporation rates we find in \S\ref{s:photevap} are of a similar order of magnitude to mass accretion rates, we suggest here that accretion can instead be modulated by coronal X-ray luminosity. We find that the bulk of the disk mass loss occurs at a radial distance of 20~AU, and it is this evaporative flow that can ``starve'' the viscous flow of gas to the inner disk.  A recent disk photoevaporation study by \citet{Gorti.Hollenbach:09} also finds the 10-20~AU region to dominate the photoevaporative flow, with mass loss rates similar to those from our study (though in their model the flow is primarily driven by FUV heating; we defer a more complete discussion of the model differences to future work).  Were all inward gas flow to be curtailed at a radial distance of 10-20~AU, the disk inward of this point would essentially be cleared of gas on the viscous time scale.  For our canonical disk model, taken from \citet{D'Alessio.etal:98}, the inner 10~AU contains only a few percent of the total disk mass and the clearing time is of order $10^5$~yr.
 
The inner disk clearing time also represents the timescale for which a given star must have a relatively high X-ray luminosity for the accretion flow to become completely starved of gas by photoevaporation.  While it is possible that shorter periods of high X-ray luminosity could give rise to depleted radial ``stripes'' that are manifest in a lower accretion rate on reaching the very inner disk, the correlation between the two properties $L_X$ and $\dot M$ could in such a case be erased during any given snapshot in time.  

The inner disk clearing time is important in the context of the large range of observed X-ray luminosities of T~Tauri stars.  \citet{Preibisch.etal:05} noted the very large scatter of observed X-ray luminosity in relation to other stellar parameters, such as stellar mass.  In that case average deviations amount to $\pm 0.7$~dex from the mean relation.  In order for photoevaporation to be at least partially responsible for the observed correlation between accretion rate and X-ray luminosity, the scatter in the latter must be due mostly to long-term intrinsic differences rather than short-term variability.  \citet{Preibisch.etal:05} ruled out stellar flaring and short-term stochastic variability as being responsible for the $L_X$ scatter, largely by comparison of COUP observations with shorter {\it Chandra} ACIS ONC observations obtained several years earlier which indicated a median deviation of only a factor of $\sim 2$.  We confirm this conclusion by comparison of stellar $L_X$ values from COUP with those from the {\it Chandra}~HRC ONC survey of \citet{Flaccomio.etal:03}: when sources of scatter due to different analysis procedures and different instrumental bandpasses are accounted for, the deviation from parity between the surveys only amounts to about a factor of 2.  Nevertheless, the $L_X$ scatter among T~Tauri stars represents a possible test of the relative importance of X-ray photoevaporation: should future studies, perhaps over
longer time baselines, find the scatter to be caused by relatively short term 
changes (activity cycles perhaps, that have not yet been seen in T~Tauri stars), then
photoevaporation-starved accretion could not give rise to the observed $L_X$--$\dot M$ correlation.

Finally, while it is suggested here that X-ray photoevaporation can drive the 
correlation between X-ray luminosity $L_{X}$  and mass accretion rate, and
consequently the X-ray luminosity differences between CTTS and WTTS, 
the true explanation might involve more than one mechanism, and it is possible 
that both coronal disruption and photoevaporation play a role.



\section{Summary}
\label{s:summary}

New calculations of T~Tauri disk photoevaporation due to stellar X-radiation find mass 
loss rates highly correlated 
with X-ray luminosity and of a similar magnitude to mass accretion rates.
The results lead us to suggest that irradiation of protoplanetary disks by 
stellar X-radiation can starve the accretion flow onto the central star 
through photoevaporation.  Such a process could explain the 
higher average X-ray luminosities of non-accreting T~Tauri stars compared with 
those of accretors.
We find from an analysis of the Ca~II~8662~\AA\ accretion 
diagnostic for T~Tauri stars in Orion that the mass accretion rate is 
also correlated with stellar X-ray luminosity, as expected in the 
photoevaporation-starved accretion scenario.

\acknowledgments

JJD was funded by NASA contract NAS8-39073 to the {\it Chandra X-ray
Center} (CXC) during the course of this research and thanks the CXC
director, Harvey Tananbaum, and the CXC science team for advice and
support.  JJD also thanks the EU ISHERPA program for financial support
during his visit to the Osservatorio Astronomico di Palermo, and the
Osservatorio director, Prof.~S.~Sciortino, and staff for their help
and warm hospitality.  EF and GM acknowledge financial support from
the Ministero dell'Universita' e della Ricerca and ASI/INAF Contract
I/023/05/0.

\bibliographystyle{apj}

\newpage

\end{document}